# THE PHYSICAL REALITY UNDERLYING THE RELATIVISTIC MECHANICS AND THE GRAVITATIONAL INTERACTION

Maurizio Michelini
*maurizio.michelini@casaccia.enea.it*
ENEA, Casella postale 2400, Rome, Italy

**Abstract** – In the present paradigm the space is filled with a very high flux of very small quanta whose wavelength equals the Planck's length. The energy $E_o = h_o \nu$ of these quanta is very small, so the relevant Planck's constant $h_o$ is much smaller than $h$. This physical paradigm imposes to the motion the principles of conservation of energy and momentum, as well as the laws of relativistic mechanics. The strong version of the equivalence principle, which states that *both* inertia and gravitation come from a *unique* phenomenon, is the *relevant test* to verify the physical reality of the cosmic quanta. Through Compton's interaction, each quantum colliding two masses gives them a little momentum $(E_o - E_1)/c$ which produces a newtonian force *pushing* the masses each towards other. The constant $G = K_o \phi_o E_o A_o^2 / 4\pi c$ depends on the quanta, so the Newton's gravitational mass no longer holds. The new gravitational force within stars depends also on the reducing quantum energy, so $G$ multiplies by the gravity factor $(n_{eq}/a) \geq 1$ depending on the star density. The highest number $(n_{eq}/a) \approx 200 \div 300$ pertains to neutron stars, which increment notably their accretion capacity. This property explains the mistery of the obscure supermassive bodies whose gravitational effects have been observed to rise up to $3.7 \times 10^6$ times the Sun effects. Current theories do not give a convincing explanation of this phenomenon.

## Introduction

According to a theoretical thought of Y. Zel'dovic, the physical interaction implying the transmission of forces to the particles is linked to the energy density of the "void" space, for which since 1967 he obtained a guess comprised between 55 and 120 orders of magnitude higher than the average mass-energy density of the universe. This very uncertain estimation did not give rise to further investigations.

On the other hand, the origin of the very powerful inertial forces is up to now not yet found.

What physical mechanism originates, within times so short to prevent any measurement, the forces which fragment a rotating steel disk ? The old attempts to correlate this force with the gravitational field of the distant masses of the universe appears today only an obsolete conjecture.

A recent work [1] proposes that space is uniformly filled with a very high flux of small quanta characterized by a wavelength equal to the Planck's length. These quanta, through Compton's collisions with matter, originate both gravitational and inertial forces, thus satisfying the strong version of the equivalence principle .

Demonstrating the properties of these small quanta presents the same task that physicists had to solve after the enuciation in 1905 of the Special relativity. At the time, the discussion about a physical explanation diverted towards the time-dilatation and the so-called time paradoxes which took many books, but gave no advance in physical science.

One of the cognitive tasks that Special relativity did not fulfil was the need of explaining the mass-velocity relationship in terms of interaction between waves and particle.

## 1 - The physical paradigm originating the relativistic mechanics

In the present paradigm the space is no longer a mathematical entity (however much sophisticated with mass-dependent geometry), but is filled by a high flux of small quanta with energy $E_o = h_o \nu_o$ and momentum $h_o \nu_o/c$, where the constant $h_o$ shows the same dimensions of the usual Planck's constant, but is enourmously smaller [1]. Firstly we notice that this paradigm endorses *entirely* the postulate of constancy of the light velocity. In fact the photon itself is constituted of cosmic quanta.

*The momentum of a particle moving within the flux of cosmic quanta*

Consider a particle with rest mass $m_o$, which moves uniformly in a straitline with velocity **v**, subjected to collisions with the isotropic flux of quanta. The particle receives (along the **v** direction) the momentum $-\Delta q_f = N_f h_o \nu_f /c$ in front and the momentum $\Delta q_b = N_b h_o \nu_b /c$ behind.

Any wave colliding with particles undergoes the Doppler effect. In the case of cosmic quanta the Doppler frequencies of quanta incident on a particle with velocity **v** are $\nu_f = \nu_o (1+ v/c)^{1/2}/(1-v/c)^{1/2}$ and $\nu_b = \nu_o (1-v/c)^{1/2}/(1+v/c)^{1/2}$, respectively in front and behind. The velocity is not referred to an external system of coordinates, but is measured (strictly speaking) through the Doppler shift of the incident quanta. The particle brings its own system of measuring its momentum/speed. This explains why the inertial forces arise without any *physical* link to external objects. The properties of the so-called inertial systems depend on the cosmic flux "anchorage" to the large masses of the universe.

Considering that the collision time of a small quantum with frequency $\nu$ is $\tau = 1/\nu$, the number of simultaneous collisions upon the particle with cross section $\sigma$ results $N_f = \sigma \phi_f \tau_f$ in front and $N_b = \sigma \phi_b \tau_b$ behind. For instance, the number of *simultaneous* collisions upon a nucleon is of the order of $10^{51}$. Betwen the flux and the particle there is a continuous equilibrium, since the particle momentum **q** equals the net momentum of the colliding quanta : $\mathbf{q} = \Delta q_b - \Delta q_f$, which substituting becomes

$$(1) \qquad |\mathbf{q}| = (h_o \sigma /c)(\phi_b - \phi_f)$$

where $\phi_f, \phi_b$ are the perturbed fluxes, in front and behind, after the interaction.

For *freely* moving particles the conservation of the unperturbed energy density $\ni_o = E_o \phi_o /c$ requires that the density *before* and *after* the interaction be equal, in front and behind

$$(2) \qquad E_o \phi_o /2c = \phi_f h_o \nu_f /c = \phi_b h_o \nu_b /c$$

where $\phi_o$ is the unperturbed total flux. Substituting the Doppler frequencies $\nu_f$ and $\nu_b$ in eq(2) and subsituting the resulting perturbed fluxes in eq(1), one obtains the momentum

$$(1a) \qquad \mathbf{q} = (\sigma E_o \phi_o /\nu_o c^2)\, \mathbf{v}/(1- v^2/c^2)^{1/2}$$

which coincides with the special relativity expression when the rest mass of the particle is defined through the cosmic quanta constants

$$(3) \qquad m_o = E_o \sigma \phi_o /\nu_o c^2 = h_o \sigma \phi_o /c^2.$$

The mass-velocity dependence $m(v) = m_o/(1- v^2/c^2)^{1/2}$ arises from the interaction of particles with the cosmic quanta. Obtaining the relativistic eq.(1a) from the wave-particle interaction (eq.1) proves that special relativity and cosmic quanta paradigm are two faces of the same phenomenon.

The fundamental principles of physics (conservation of energy and momentum) appear to be conditions imposed by the cosmic quanta to the motion of masses.

*The constants characterizing the cosmic quanta*

From eq.(3) we have the corollary that the cross section of each particle is proportional to its mass through the constant [1]

$$(4) \qquad A_o = \sigma /m_o = c^2 / h_o \phi_o \approx 4.63 \times 10^{-11}\ \mathrm{m^2/Kg}.$$

This relationship recognizes that the inertial mass originates from the interaction of particles with the quanta. From the knowledge of the constant $A_o$, from the expression of $G = K_o \phi_o E_o A_o^2 / 4\pi c$ in terms of the cosmic quanta (see eq.9) and from a condition on the cosmic quanta pressure [1], we obtained the principal constants characterizing the quanta (SI system)

$$\lambda_o = 4.05 \times 10^{-35} \qquad E_o \approx 5.91 \times 10^{-61} \qquad э_o \approx 4.79 \times 10^{61}$$
$$\phi_o \approx 2.43 \times 10^{130} \qquad h_o \approx 7.86 \times 10^{-101} \qquad \nu_o \approx 7.41 \times 10^{42}.$$

Differently from photons, which are quantized waves of the electromagnetic field, the cosmic quantum waves originate the inertial and the gravitational interaction.

Photons are generated by the accelerated motion of electric charges. On the contrary the cosmic quanta pre-exist to the masses. Both types of quanta are characterized by the undulatory properties.

From a quick inspection of the above constants one could realize that such a flux makes crowded the space we know. The density of the cosmic quanta is very high $(\phi_o/c) = 8.1 \times 10^{122}/m^3$. One may suspect that so many quanta could interfer each other, originating chaotic turbulence instead of ruling the motion of masses. To this subject it has been recognized [1] that the mutual cross section of the cosmic quanta is so small that the mean free path in the space is of the order of $10^{21}$ m. Hence the local crowd is only apparent.

Also photons have a very small mutual cross section, so they move freely through the whole space.

The energy density of the cosmic quanta $э_o \approx 4.79 \times 10^{61}$ joule/m$^3$ is very high, but this is not a problem because the powerful inertial forces are generated by particle interaction with the cosmic flux. Hence the local energy reservoir must be very intensive. This explains why the inertial forces arise without any appreciable delay respect to the acceleration of masses.

In the frame of the quantic description of the inertial interaction, eq.(1) defines the momentum of a particle. The consequent eq(1a) is formally equal to the relativistic definition, proving that between Special relativity and cosmic quanta there is a tight correspondence. However the last paradigm furnishes much more informations. For instance eq.(3) shows the structure of the rest mass which looks like an assembly of an enormous numbers of quanta with energy $E_o$, simultaneously colliding upon the cross section $\sigma$.

This gives an indication about the origin of the *new* particles appearing in the high energy collisions.

*The inertial force in the uniform straight motion*

The inertial forces can be derived directly through the quantic description of the interaction.

For instance the force acting on a *free* particle with a uniform velocity **v** can be obtained from the difference between the forwards force $f_f = \sigma \phi_f \Delta q_f$ and the backward force $f_b = \sigma \phi_b \Delta q_b$ generated from the collisions with quanta

$$(4) \qquad f = \sigma(\phi_f \Delta q_f - \phi_b \Delta q_b) = (h_o \sigma / c)(\phi_f \nu_f - \phi_b \nu_b).$$

This inertial force is identically null as descends from eq.(2) which imposes the conservation of the quanta energy density.

*The inertial force in the accelerated straight motion*

Upon an *accelerated* particle wich moves in a straight line the inertial force is

$$(5) \qquad f = (h_o \sigma / c)(\phi_f \upsilon_f - \phi_b \upsilon_b)$$

where $\upsilon_f, \upsilon_b$ are the quantum frequencies in presence of acceleration.

Since the particle undergoes a velocity variation $\Delta v$ during the collision time, the wavelength of the incident quanta must consider, besides the Doppler effect, the acceleration of the particle. Hence the wavelength in front is $\lambda_f = \lambda_f - \tau_f \Delta v = \lambda_f(1 - \Delta v/c)$ and behind $\lambda_b = \lambda_b - \tau_b \Delta v = \lambda_b(1 + \Delta v/c)$
where $\lambda_f, \lambda_b$ are the Doppler wavelengths.

From these relations we obtain the frequency of quanta incident in front $\upsilon_f = c/\lambda_f$ and behind $\upsilon_b = c/\lambda_b$. Substituting in eq.(5) and taking into account eq.(4) one gets

(5a) $\qquad f = (h_o \sigma \Delta \mathbf{v}/c^2)(\phi_f \nu_f + \phi_b \nu_b)$.

In presence of forces acting on the particle, the conservation of the energy density becomes simply
$$\mathfrak{z} = \phi_f h_o \nu_f/c = \phi_b h_o \nu_b/c.$$
Substituting in the preceding equation and recalling the definition of the rest mass, we obtain the finite difference equation
$$f = (\phi_f \nu_f + \phi_b \nu_b) m_o \Delta \mathbf{v}/\phi_o = (2\phi_f \nu_f/\phi_o) m_o \Delta \mathbf{v}.$$
Substituting the Doppler frequencies $\nu_f, \nu_b$ one gets the quantic inertial force in the straightline motion in function (for instance) of the flux $\phi_f$

(5b) $\qquad f = (2\phi_f/\phi_o)(1 + v/c) m_o \Delta \mathbf{v}/\tau_o(1 - v^2/c^2)^{½}$.

To explicit this expression it is necessary to make recourse to the balance between *momentum* and *impulse* within the time of interaction

(6) $\qquad f \tau_o = \Delta \mathbf{q}$

which recalling eq(1a) becomes

(6a) $\qquad f = - m_o (\mathbf{v} + \Delta \mathbf{v})/\tau_o[1 - (v + \Delta v)^2/c^2]^{½} + m_o \mathbf{v}/\tau_o[1 - v^2/c^2]^{½}$.

In the straight motion this gives the quantic inertial force

(5c) $\qquad f = - m_o \Delta \mathbf{v}/\tau_o[1 - v^2/c^2]^{3/2}$.

Comparing with eq(5b) we obtain the forwards flux
$$\phi_f/\phi_o = 1/2(1 + v/c)(1 - v^2/c^2)$$
and the backward flux
$$\phi_b/\phi_o = 1/2(1 - v/c)(1 - v^2/c^2).$$

To see the interaction from the *continuum* standpoint, we may substitute $\Delta v \cong \tau_o \, dv/dt$, thus obtaining the relativistic formula

(5d) $\qquad f \cong - m_o \, dv/dt \, [1 - v^2/c^2]^{3/2}$.

The two descriptions, eqs(5c,5d), are equivalent for any practical purpose, since $\tau_o \approx 10^{-43}$.
The small difference betwen the quantic descriptions and the relativistic definitions shows that the inertial forces are not generated from the void space (*continuum*), but from the interaction of particles with the flux of cosmic quanta.

*The centrifugal force*

Analogously, the quantic description of the circular centrifugal force is obtained putting in eq.(6a) $\Delta v = 0$, so obtaining

(7) $\qquad \mathbf{f}_c = - m_o \Delta \mathbf{v}/\tau_o [1 - v^2/c^2]^{1/2}$.

This reduces to the relativistic expression putting $\Delta \mathbf{v}/\tau_o = \omega^2 \mathbf{r}$

(7a) $\qquad \mathbf{f}_c = - m_o \omega^2 \mathbf{r}/[1 - v^2/c^2]^{1/2}$.

**2- Critique of the concept of gravitational mass**

The logical-mathematical construction of General relativity was conceived to describe accurately, in the frame of the relativistic mechanics, the astronomical observations - carried by photons with finite velocity - of the gravitational phenomena. Like the newtonian gravitation, general relativity postulates the *universal* constant *G*.

The success found by General relativity within the last 80 years descends from this large theoretical basis. Many people think that such a profound theory of the observations cannot be nothing else than a theory of gravitation. But this belief does not have a scientific basis.

The right question is: Can general relativity be considered a physical theory of the gravitational force, in contrast with the modern view requiring an interaction between matter and *field-carrying waves* to originate any force? The ample experimental research of gravitational waves did not obtain a positive answer. On the contrary there have been theoretical works [2] claiming that gravitational waves cannot be produced by motion of bodies.

Of course, in the present paradigm the gravitational force is just due to waves, i.e. the flux of cosmic quanta. But these very small waves are not the searched "macroscopic" waves generated by acceleration of macroscopic stellar masses.

Besides, some epistemologists are not convinced that general relativity is a gravitational theory for a basic reason. The strong version of the equivalence principle requires that inertia and gravitation originate both from *the same* physical phenomenon.

According to general relativity this phenomenon resides in the property of the spacetime to bend under the presence of masses, thus generating the *gravitational* forces. On the contrary, the *inertial* forces are produced by a simple *flat* space, which is an evident generalization of the Newton's *spatium absolutum*. However Newton correctly admitted not to possess the *physical* key of this concept (*Hypotheses non fingo*). The empty space is not the underlying reality that generates both the inertial and the gravitational forces.

As we shall see on paragraph 4, the cosmic quanta paradigm gives rise to a newtonian force which *pushes* the masses each towards other, thus satisfying the equivalence principle.

Moreover the recent observations of the galactic supermassive obscure bodies, which did not find a consistent explanation within the classical/relativistic gravitation (paragr. 7,8,9), induce to think that supermasses can be explained when the constant *G* is not universal, but depends on the mass and density of each celestial body.

### 3- Cosmological observations and general relativity

We concisely resume the present situation about the cosmological model, although we believe that the final test about the gravitation theory will not come from cosmology. The recent observations through the Hubble orbiting space telescope pointed out some discrepancies respect to the predictions of the relativistic model of the universe with flat spacetime, which predicts a mean density of matter

$$\delta_{cr} = 3H^2/8\pi G.$$

Adopting the average $H$ from the Hubble observations [3], $\delta_{cr}$ results about 30 times higher than the density of the luminous matter observed by the astronomers. The discrepancy reduces to 3÷4 times considering the invisible matter estimated by means of its gravitational effects upon the neighbouring luminous matter.

The *missing mass* problem can hardly be solved hypothesizing much more invisible matter. Since the antimatter predicted by the *Big bang* model has not been found in our universe, it was proposed the existence of a non barionic *obscure* matter deprived of radiative emission, but with gravitational effects.

A natural candidate to this role are the neutrinos, which fill the universe with a density of about $1.2 \times 10^8$ /m$^3$. A recent work of A.Melchiorri e R.Trotta [4] has pointed out that the neutrino energy estimated by means of the Superkamiokande experiment appears too little to explain the *missing* mass. Further experimental proofs are in progress. In any case we observe that the flux of neutrinos is too low to originate (in the frame of the present paradigm) the inertial forces through their small cross section with matter.

It has been also proposed, with little success, an *obscure* energy fulfilling the space.

According to L.Krauss and M.Turner [5] the "*The cosmic puzzle*" may be explained reintroducing in the relativistic models a cosmological constant, initially devised by A.Einstein in 1916 to study a static universe and subsequently missed when the expanding models came out.

In practice, this solution lost its interest in comparison with the exciting new cosmological observations suggested by L.Krauss itself. To refine the measurements of the Hubble constant, Krauss proposed to measure the large cosmological distances by means of the supernovae Ia, which act in the universe as "standard candles" being their luminosity tendentially constant everywhere.

These measurements, interpreted in the expansion optique, show that the universe is not decelerating, due to the gravitational force, but is accelerating [5] contrarily to the paradigm of gravitational-mass.

The discussion about the interpretation of this result is in progress.
T.Davis and C.Lineweaver [6] pointed out the present "expanding confusion" in the cosmological problem, in contrast with the clarity of the first relativistic models.

In any case it seems doubtful that a firm conclusion about the gravitation theory may come from cosmology.

## 4- Generation of the gravitational force through the cosmic quanta

As previously stated, the equivalence principle requires that gravitation must be generated by the same phenomenon which originates the inertial forces. This principle furnishes the main test to verify the physical reality of the cosmic quanta.

It is known that single photons colliding with matter through Compton's effect, give up some little energy/momemtum to the particle. Extending the Compton's equations to the cosmic quanta interaction with matter, the origin of the gravitational force has been shown in detail [1].

Here we show the same result with a very simple procedure, which nevertheless catches the essence of the phenomenon [7] . The process of transmission of a little momentum from the cosmic quanta to the masses has the appearance of *continuity,* since the number of cosmic quanta colliding simultaneously on a nucleon is of the order of $10^{51}$. Before to collide, these quanta have done very different pathes. Those which made the longest path (which is of the order of the m.f.p. $10^{21}$m) possess the oldest energy $E_o$ which is greater than the energy $E_1 = E_o - \Delta E$ possessed by quanta coming from a collision with a local particle, being $\Delta E$ the small energy given up to that particle.

Let's consider a nucleon of mass $m$ placed at a distance $r$ from a stellar mass $M$ which is *opaque* to the quanta, since the sum $\sigma M/m$ of the cross sections of all nucleons (and other particles) is much greater than the geometrical cross section $\pi R^2$ of the mass $M$, thus implying a complete screening of the cosmic quanta. The force which pushes the particle towards the mass $M$ depends on the difference between the momentum $\Delta q_n = E_n/c$ given up by the beam $\gamma(r)\phi_o$ of weakened quanta coming from the interior of the star, being $\gamma(r) = \pi R^2/4\pi r^2$ the solid angle subtended by the star, and the momentum $\Delta q_o = E_o/c$ given up by the opposed equal beam of quanta coming from the external space.
Then the centripetal force upon the particle results

$$(7) \qquad f(r, E_n) = \sigma\, \gamma(r)\phi_o\, (E_o - E_n)/c \ .$$

We shall call "gravitational" this force although in reality the two masses do not *draw* each other, but are *pushed* each towards the other. This is the reason why the old paradigm of gravitational-massis ruled out.. Due to the *n* Compton collisions suffered on the average at the interior of the mass, the energy of the coming out quanta reduces to[1]

$$(8) \qquad E_n = E_o/(1+ nK_o),$$

where $K_o = E_o/mc^2 \approx 4\times 10^{-51}$, so the force becomes

$$f(r,n) \cong n\,(\sigma\, K_o \phi_o\, E_o)\, R^2 /4cr^2$$

being $nK_o \ll 1$ for all stars. The small momentum the colliding quanta give up is $\Delta E/c = K_o E_o/c$.

Putting $l_o = m/\sigma\delta = 1/A_o\delta$ the quanta m.f.p. within the mass $M$ of density $\delta$, let's introduce the mass optical thickness $a = (4/3)R/l_o = A_o M/\pi R^2$. Substituting $a$ in the preceding equation and defining the product of the quanta constants

$$(9) \qquad [K_o \phi_o\, E_o A_o^2/4\pi c] = G.$$

one gets the almost familiar relationship

$$(10) \qquad f(r,n) \cong (n/a)\, G\, M m/r^2$$

which differs from the newtonian expression by the term $(n/a)$.

The ratio $(n/a)$ is called the *gravity factor* of the mass. It multiplies the newtonian force to take into account the energy reduction undergone by quanta in the repeated collisions within the mass.

It can be proved that $(n/a) \geq 1$. In fact the optical thickness $a$ represents the average number of m.f.p. travelled by a (hypothetical) quantum moving in a straight line, whereas $n$ is the average number of real collisions (i.e. the number of travelled m.f.p.), each producing a deviation of the quantum. If the mass is sufficiently dense, the quantum zigzags long time before coming out with a reduced energy $E_n$, so that $(n/a) > 1$. If the mass is *transparent* to the cosmic quanta, that is $a = A_o M/\pi R^2 \ll 1$, then one obtains $(n/a) = 1$.

For instance the Earth optical thickness is $a \approx 2.2$, the Sun shows $a \approx 65$, white dwarfs show $a \approx 10^5$.

## 5- The gravity factor of very dense masses

As long as the optical thickness does not exceed a threshold (for instance $a \approx 10^5$), the result $n \cong a$ holds. It has been shown [1] that the gravitational force defined by eq.(10) coincides with the newtonian force as long as stars and planets are constituted of neutral or weakly ionized atoms whose pressure is given by the ideal gas equation. The corresponding gravity factor $(n_{eq}/a) \cong 1$ confirms the accuracy of the newtonian gravitation for the ordinary celestial bodies.

The supergravity, i.e. the rise of the gravity factor $(n_{eq}/a) \gg 1$, originates from the occurrence of the weakened quanta $E_n$ coming out from a very dense star after numerous collisions. In this way the gravitational force depends also on the quanta energy. The average number of collisions $n_{eq}$ corresponding to the stability of the star can be calculated by imposing the equilibrium between the gravitational pressure and the pressure of gas and radiation.

In the middle of massive stars (tens of solar masses) the plasma is constituted entirely of charged particle (electrons and nuclei). The pressure of a gas made of particles with average kinetic energy $\ni$ is defined as the energy density per unit of volume

$$(11) \qquad p = \ni (\delta/m).$$

In the degerate matter the kinetic energy of an electron $m_e v^2/2$, which produces the centrifugal force $\hbar^2/m_e r^3$ balancing the electrical force, takes the average energy

$$\ni_e = m_e v^2/2 = \hbar^2/2m_e r^2$$

where $r$ is the nucleus-electron distance. At the critical density $\delta_{cr} \approx 3\times 10^{16}$ the nuclei merge, so protons and electrons interact *individually*. In this case the proton-electron distance takes the average $r \cong \Delta x/2$ depending on the average distance $\Delta x$ between two nucleons within a plasma of density $\delta = m/\Delta x^3$.

Assuming the parity of electrical charges, one obtains the pressure of degenerate matter which is present, for instance, in the core of massive stars

$$p_{de} \cong (\pi^2 \hbar^2/5 m_e m^{5/3}) \delta^{5/3}.$$

The core stability is guaranteed by the equilibrium between the gas pressure and the gravitational pressure (assuming the radiation pressure can be neglected)

(12) $\qquad p_{de} \cong 1.05 \times 10^7 \delta^{5/3} = p_{gr} \cong 0.459\, n_{eq}\, \delta^{2/3}\, M^{1/3}.$

where $\delta$ is the average density. Isolating $n_{eq}$ and dividing by the optical thickness $a = A_o M/\pi R^2$, one gets the gravity factor

(13) $\qquad (n_{eq}/a)_{de} \cong 5.8 \times 10^{17}\, \delta^{1/3}/M^{2/3}$

of dense celestial masses such as the massive stars as well as the white dwarfs.

Since the white dwarfs show a mass not exceeding 1.4 solar masses and an average density of about $10^9$ kg/m$^3$, the gravity factor results around unity, so the gravitational force does not differ sensibly from the newtonian force.

Eq.(13) applies also to the dense (average $\delta \approx 10^{12}$) massive stars with $M \approx 30 \div 50$ solar masses, giving a gravity factor around unity.

## 6 - The origin of the neutron stars and the supernova sudden collapse

It is known that in the central region of stars with more than a few tens of Sun masses, the increasing density triggers the electron capture by protons, thus originating a core almost entirely made of neutrons with average energy $\ni_n$ derived from the captured electron. After the star collapse and consequent explosion (*supernova*), a neutron star appears in place of the progenitor star.

The pressure of the high density neutron gas (which seems to behave like superfluid) resulted [8] $p_n = \ni_n (\delta/m) \approx 10^{18} \delta$, which repeating the procedure of pressure balance (eq12) gives the gravity factor

(14) $\qquad (n_{eq}/a)_n \cong 9 \times 10^{28} R_n / M_n.$

In the case of isolated neutron stars with radii $R_n \approx (1 \div 3) \times 10^4$ m. and mass $M_n \approx 2 \div 5$ solar masses [9] the gravity factor results in the range $(n_{eq}/a)_n \approx 206 \div 305$.

High gravity factors are typical of neutron stars because they form in a very rapid way through the supernova explosion, which "freezes" the neutron gas characteristics.

The shortening of the nuclear reaction cycles, which maintain within stars the high temperature and pressure, produces a slow star contraction. The increasing density favours the electron capture, so the neutron fraction rises within the core. Obviously, the core mass does not change, as well as its newtonian gravitational force, so the contraction proceeds slowly.

What triggers the sudden star collapse, which happens within *some seconds,* according to the work of A.Fruchter, A.Levan, et al. [10]?

Within the present paradigm the cause is the growing core gravity factor due to the growing neutron density during the star collapse. Let's compare the gravity factor (eq.13) of the core degenerate matter

(13a) $\qquad (n_{eq}/a)_c \cong 5.8 \times 10^{17}\, \delta_c^{1/3}/M_c^{2/3}$

where $\delta_c \approx 10^{14}$ is the core density before the neutron growth, with the gravity factor of the neutron core (eq.14) put in the form

(14a) $\qquad (n_{eq}/a)_n \cong 5.59 \times 10^{28}/\delta_n^{1/3} M_n^{2/3}$

where the density is assumed equal to the neutron star density $\delta_n \approx 10^{17}$. Since during the electron capture the core mass does not change ($M_c = M_n$), substituting the densities, one gets the ratio

(15a) $\qquad (n_{eq}/a)_n / (n_{eq}/a)_c \approx 4 \div 5$

showing that the core gravitational force enhances considerably when the residual protons become neutrons through an avalanche electron capture that happens within some seconds.
This fact may be responsible of the *sudden* collapse involving large masses, which has no clear origin in the classical/relativistic gravitation. Without a jump in the core gravitational force, the star would not suddenly collapse and the consequent bounce with the *supernova* explosion would not happen.

**7 -The observation of supermassive galactic obscure bodies**

The crucial test for the gravitational theory will probably come from the recent observations of supermassive obscure bodies discovered within the active galactic nuclei (AGN).
In 2002 R.Schoedel et al.[11] discovered, after decennial observation of an orbiting star, in the middle of the Milky Way a pointy obscure object whose gravitational effects on the neighbouring stars are those of a body with 3.7 million solar masses.
In 2004 in the same galactic region a minor obscure object was found with "only" 1300 solar masses.
Finally, in January 2005, L.Miller [12] communicated to the Conference of the American Astronomical Society the discovery, thanks to the lucky observation of three orbiting luminous bodies, of an obscure object with gravity equivalent to 300.000 solar masses placed in a galaxy 170 million light years away. This type of research promises to reveal a consistent number of supermasses.
It is known that obscure bodies with masses between 3 and 21 solar masses have been observed in numerous systems of binary stars. The greatest of these masses formed by gravitational accretion upon neutron stars swallowing gaseous mass from the luminous companion. Neutron stars originate from the explosion (*supernova*) of massive stars whose masses do not exceed those ascertained within the globular stellar clusters, where the probability to find a star with mass greater than 150 solar masses (up to now never observed) has been calculated equal to $10^{-8}$ by D.Figer [13]. The famous Plaskett's binary, for instance, shows a mass not exceeding about 75 solar masses.
In this frame, the observation of the obscure supermassive bodies generated a profound uncertainty. Although the hypothesis appears up to now speculative, the observed supermasses might come from *supernovae* of primordial supermassive stars.
R.Larson and V.Bromm [14] in a work searching for the formation of the first stars (the so called Population III) found through simulations a typical star mass around 100 Sun masses, with a rough upper limit of 700 Sun masses. However these astronomers do not formulate the hypothesis that obscure supermasses may be the remnant of ancient very massive supernovae. They recall also the possibility, unique to the zero-metallicity first stars, of the complete disruption of the progenitor in a supernova leaving no neutron star. A preceding work by T.Abel, G.Bryan and M.Norman did not find more than 300 solar masses. Substantially the possibility that obscure supermasses come from supernovae of ancient supermassive stars appears unlikely.
On the other hand, possible adjustments of the gravitational theory appear not adequate. It is necessary to revise the physical basis of gravitation.
Before to examine the possibility that obscure supermasses may depend on the large gravity factor of neutron stars, we have to discuss the possibility, in the frame of classical gravitation, of mass accumulation upon a neutron star through the gravitational accretion of galactic gas.

**8 -Current hypothesis on the formation of obscure supermasses**

T.Heckman, G.Kauffmann et al.[15] studied the formation of great obscure masses by beans of gravitational accretion upon neutron stars. The accretion rates are estimated around $10^{-3}$ solar masses per year when the gravity of the obscure body is equivalent to $3 \times 10^7$ solar masses.

This grow rate is not much different from that observed for neutron stars in binary systems (with masses between 3 and 21 Sun masses) where the mass accretion has been assessed around $10^{-8}$ Sun masses per year through observation of X-ray emitted by the accreting disk of gas swallowed from the luminous companion. This mechanism is powerful between two stars at a distance around $10^{10}$ m, but the extension to obscure galactic supermasses shows little credibility due to the large distances (around $10^{16}$ m) between the obscure body and the nearest stars to be swallowed.

It appears more reliable to consider the gravitational accretion of galactic gas.

Let's consider a star with mass equal to the critical Jeans mass for a primordial cloud of hydrogen. After the radiation blew the interstellar gas, this massive star collapsed in *supernova*, leaving a neutron star immersed in a low density cloud.

Current scenarios of a neutron star placed within a high density cloud of an active galactic nucleus (AGN) has little significance. In any case the probability that an accreting neutron star might draw large masses of interstellar gas *in competition* with the formation of ordinary stars results very poor in the work by M.Krumholz et al.[16].

Also the work of T.Heckman, G.Kauffmann et al. shows that in AGN galaxies the mass grow rate of new stars is about $10^3$ times the grow rate of obscure bodies by accretion .

In the primordial scenario the massive neutron star within a low density cloud (where the formation of stars by gravitational collapse does not take place) could draw the whole cloud mass.

T.Heckman, G.Kauffmann et al.[15] affirmed that, in the past, supermasses could have grown up to $10^7 \div 10^8$ Sun masses *provided* they had sufficient time to draw the galactic gas.

Thus the problem is to verify in the frame of classical gravitation how much mass the obscure bodies may accumulate during the time allowed by the standard age of the universe.

**9 - A model to calculate the potentiality of galactic gas accretion in classical gravitation**

Our aim is to calculate, within the frame of classical/relativistic gravitation, the time that an isolated neutron star, placed in the middle of a large cloud of primordial hydrogen, requires to accumulate mass up to $10^7 \div 10^8$ solar masses. In other words we want to establish the *maximum potentiality* of gravitational accretion without occurence of adverse conditions (presence of cloud angular momentum, magnetic fields, ecc.).

Assuming the standard age (14 billion years) of the universe, the duration of the star accretion does not probably exceed 8-9 billion years, since we must take into account the time of formation of the primordial galaxy and the time of occurence of a supernova which produced a massive neutron star.

The gravitational accretion upon a neutron star immersed in a spherical cloud is given by the mass of gas whith density $\delta(r,t)$ transiting with net radial velocity $u(r,t)$ through a spherical surface at a distance $r$ from the obscure star of mas $M$

(15) $\quad dM/dt = 4\pi r^2 u(r,t) \delta(r,t)$.

Let's fix our attention on the stationary situation where all the quantities in eq.(15) attained spatial stable distributions $\delta(r)$ and $u(r)$. The radial speed is much reduced respect to the free falling velocity because of the molecule collisions , which increase when the gas density increases.

Putting the average time of flight $\tau(r) = l_g(r)/v_g = (m/\sigma)_g/v_g \, \delta(r)$, where $l_g(r)$ is the molecular mean free path (assumed to be $l_g(r) \ll r$), one finds the radial velocity

(16) $\quad u(r) \cong (GM/r^2)\tau(r) = (GM/r^2)(m/\sigma)_g/v_g \, \delta(r)$

where $\sigma_g$ is the molecule cross section and $v_g = (2kT/m)^{1/2}$ is the molecular speed related to the primordial gas temperature $T$, assessed around 200-300°K [14]. Being the gas temperature little influenced by the speed of cloud contraction, $v_g$ tends to remain uniform during the travel. The radial velocity $u(r)$ holds from the outer region of the cloud up to the region ajacent to the gravitant mass $M$, where the density is so high that $u(r)$ tends to zero. At the distance $r_x$ (some radii of the mass) where takes place the production of X-ray, the gas velocity approaches the light velocity, so eq(16) furnishes indications on the local gas density. Substituting the quantities in eq(15), one finds the accretion rate towards the obscure object

$$\text{(17)} \qquad dM/dt \cong 4\pi (m/\sigma)_g GM/(2kT/m)^{1/2}$$

which does not depend on $r$, but on the gravitant mass and the gas temperature.

This corresponds to the fact that in stationary conditions the gas flow rate does not change along the distance, so it can be calculated even far from the obscure mass.

Obviously the continous effusion of gas from the cloud reduces its density $\delta(r,t)$, which may be put equal to

$$\delta(r,t) \approx \delta(r)\, [M_g - M(t)]/M_g$$

where $M_g$ is the initial cloud mass. Under these conditions eq(17) becomes

$$\text{(17a)} \qquad dM/dt \approx 4\pi (m/\sigma)_g GM (1 - M/M_g)/(2kT/m)^{1/2}.$$

As a consequence the mass accumulated by the neutron star with mass $M_o$ during the time $\Delta t$ results

$$\text{(18)} \qquad M(t)/M_g \approx M_o \exp(B\Delta t)/[M_g + M_o \exp(B\Delta t)]$$

where $B = 4\pi (m/\sigma)_g G/(2kT/m)^{1/2}$. From this solution it appears that the accretion would accumulate 90% of the cloud mass (i.e. $M/M_g = 0.9$) when the duration equals

$$\text{(19)} \qquad \Delta t_{90} = [(2kT/m)^{1/2}/4\pi G(m/\sigma)_g]\, \ln(9M_g/M_o).$$

Firstly we note that the duration does not depend very much on the initial mass $M_o$. Then the problem of a primordial neutron star mass $M_o$ of the order of some hundred Sun masses instead of some solar masses, shows no interest since the duration changes only by a factor 2-3.

Let's calculate the time $\Delta t_{90}$ for a primordial large cloud with $10^8$ Sun masses. Assuming the gas is molecular hydrogen, corresponding to the ratio $(m/\sigma)_g \approx 10^{-7}$, the duration of accreting a whole cloud of $10^8$ Sun masses upon a neutron star with $M_o = 10$ Sun masses, results $\Delta t_{90} \approx 1.4 \times 10^{12}$ years, that is 100 times the standard age of the universe !

If the cloud mass attained only $10^4$ Sun masses, the duration reduces to $\Delta t_{90} \approx 7.3 \times 10^{11}$ years, that is 52 times the standard age of the universe! It appears that the main observed supermasses did not form by accretion because the shortness of the age of the universe.

These results agree with the qualitative predictions on the long accretion time of large supermassive bodies presented in the work by T.Heckman, G.Kauffmann et al.[15] .

The primordial hydrogen galaxies did not normally contain cosmic dust. However, assuming during the accretion that a small fraction of the cloud generated powders, which takes approximatively a ratio $(m/\sigma)_g \approx 10^{-5}$, the duration of the powder accretion obviously reduced by 100 times.

However the accelerated dust accretion would marginally drag the accretion of hydrogen, which constitutes the bulk of the cloud.

In conclusion, there is no possibility in the frame of the classical/relativistic gravitation to build large obscure supermasses by accretion of galactic gas within 14 billions years.

Nevertheless the supermasses have been observed. It appears necessary to find an explanation which probably must leave the classical/relativistic gravitation.

## 10 - The evolution of the neutron stars by gravitational accretion

Differently from the white dwarfs, which do not generate appreciable accretion of interstellar gas, the neutron stars may determine large gravitational accretion of gas. This may be responsible of the observed obscure supermasses, which up to now did not find a rational explanation within the classical/relativistic gravitation theory.

In the paradigm of the cosmic quanta the key resides in the high gravity factor (eq.14) of the neutron stars

$$(n_{eq}/a)_n = 9 \times 10^{28} R_n / M_n \approx 206 \div 305$$

which obviously makes shorter the duration of accreting a certain mass.

The observed gravitational supermasses $\Psi = GM$, actually are supergravities

$$\Psi = (n_{eq}/a)_x G M_x$$

where $(n_{eq}/a)_x$ is the gravity factor of the supermass $M_x$. Hence the mass which has to be accumulated is only $M_x = \Psi / G(n_{eq}/a)_x$.

How the factor $(n_{eq}/a)_x$ changes during the mass accumulation is not clear.

We may envisage two strategies. The first one suggests that the accreted mass takes in a short time the high density $\delta_n \approx 10^{17}$ of the neutron star, so the relevant gravity factor is

$$(20) \qquad (n_{eq}/a)_x \cong 9 \times 10^{28} R_x / M_x = 5.59 \times 10^{28} / \delta_x^{1/3} M_x^{2/3}.$$

The increase of $M_x$ when the density does not change, reduces the gravity factor. Since any gravity factor is $\geq 1$, it follows that when $(n_{eq}/a)_x \cong 1$, the mass $M_x$ may rise up to $1.5 \times 10^4$ Sun masses. From this point on, the accretion continues with the classical grow rate.

The second strategy suggests that the accumulated mass requires long time to grow dense, so the large part of the star remains degenerate matter with an average density like white dwarfs.

Under these conditions we assume that the accumulated mass takes the gravity factor of the white dwarfs (eq.13), whereas the neutron core continue to follow the neutron star factor (eq.14). Summing the gravities of the two masses, one gets (when $M_x >> M_n$)

$$(21) \qquad \Psi \approx (n_{eq}/a)_n M_n + M_x (n_{eq}/a)_{de}.$$

The neutron core contracts due to the gravitational pressure of the accreting mass, so the neutron gravity becomes negligible when the observed supermass $\Psi$ exceeds about $10^5$ Sun masses.

From eq.(21) one substantially obtains that the gravity $\Psi$ of large supermasses may be expressed through the gravity of the degenerate matter $\Psi \approx (n_{eq}/a)_{de} M_x$ where the gravity factor

$$(22) \qquad (n_{eq}/a)_{de} = 5.8 \times 10^{17} \delta_x^{1/3} / M_x^{2/3}$$

reduces when the mass increases, because the density of degenerate matter attains the highest number preceding the formation of the neutron dense fluid. So, when $(n_{eq}/a)_{de} = 1$, the mass takes the value $M_x \approx 2 \times 10^{35} = 10^5$ Sun masses. Successively, the accretion continues with the classical grow rate.

In general, it is possible to devise adequate strategies such that the accretion upon a neutron star may rise up, for instance, to $10^8$ Sun masses.

The relevant problem is now to verify the possibility that accumulation of $3.7 \times 10^6$ Sun masses (i.e. the largest supermass observed) occurs *within* the standard age of the universe.

## References

1 – M.Michelini, "*The cosmic quanta paradigm fulfils the relativistic mechanics, improves the gravitation theory and originates the nuclear force*", arXiv: physics/0509017, (2005)

2 – A.Loinger , "*No motions of bodies produce GW's*", arXiv:physics/0606019 v1, 2 Jun 2006

3 – W.L.Friedman, B.Madore et al."*Final results from the Hubble Space Telescope Key Project to measure the Hubble constant*" Astrophysical Journal, **553,** 47-72 (2001)

4 – A.Melchiorri, R.Trotta , "*L'universo dei neutrini*", Le Scienze, **448,** 58-65 (Dicembre 2005)

5 – L.Krauss, M.Turner "*The cosmic puzzle*", Scientific American, Nov.2004

6 - T. Davis, C. Lineweaver, "*Expanding confusion: common misconceptions of cosmological horizons..*" *arXiv*: astro-ph/0310808, (Nov.2003)

7 – M.Michelini, "*A fundamental test for physics: the galactic supermassive obscure bodies*" arXiv- physics/0509097 (Sept.2005)

8 – M.Michelini, "*The sudden supernova collapse and explosion*" (to be published)

9 – J.Lattimer, "*Isolated neutron star*" Proceed. Conference, London, 2006, April 24-28

10 – A.Fruchter, A.Levan, L.Strolger et al., Nature, Vol. **441**, N.7092, 463-467 (2006)

11 - R.Schoedel, T.Ott et al."*A star in a 15.2 years orbit around a supermassive black hole at the centre of the Milky Way*", Nature, **419,** 694-696 (2002)

12 – L.Miller et al., Communicat. Amer. Astron. Soc. Conference, *Nature News,* (January 2005) DOI 10.1038/050110-6

13 – D.Figer, "*An upper limit to the masses of stars*", Nature*,* **434**, 192-194 (2005)

14 – R.Larson, V.Bromm, "*The first stars*", arXiv:astro-ph/0311019 v1 (3 Nov 2003)

15 – T.Heckman, G.Kauffmann, et al."*Present-day growth of black holes and bulges: the Sloan digital Sky Service Perspective*, Astron. Jour. **613**,109-118

16 – M.Krumholz, C. Mackee, R. Klein "*The formation of stars by gravitational collapse rather than competitive accretion*," Nature **438,** 332-334